\documentclass[reprint,superscriptaddress,amsmath,amssymb,
aps,prb,]{revtex4-2}
\usepackage[english]{babel}
\usepackage{amsmath}
\usepackage{amssymb}
\usepackage[bookmarks=true,colorlinks,citecolor=blue,linkcolor=blue,allcolors=blue]{hyperref}
\usepackage[capitalise]{cleveref}
\usepackage{physics}
\usepackage[toc,page]{appendix}
\usepackage{graphicx}
\usepackage[switch]{lineno}
\usepackage{orcidlink}
\usepackage{xurl}
\usepackage{bm}
\usepackage{xcolor}

\newcommand{\J}{\boldsymbol{\gamma}}
\newcommand{\Sv}{\hat{\bm{S}}}
\newcommand{\Rv}{\mathbf{R}}
\newcommand{\deltav}{\boldsymbol{\delta}}

\begin{document}

\title{Designing Strongly Correlated Quantum Phases of Matter \\ with Foundation Neural-Network Quantum States}

\author{Alessandro Sinibaldi \orcidlink{0009-0009-8498-6068}}
\email{alessandro.sinibaldi@epfl.ch}
\affiliation{Institute of Physics, \'{E}cole Polytechnique F\'{e}d\'{e}rale de Lausanne (EPFL), CH-1015 Lausanne, Switzerland}
\affiliation{Center for Quantum Science and Engineering, EPFL, Lausanne, Switzerland}

\author{Luciano Loris Viteritti \orcidlink{0009-0004-2332-7943}}
\email{luciano.viteritti@epfl.ch}
\affiliation{Institute of Physics, \'{E}cole Polytechnique F\'{e}d\'{e}rale de Lausanne (EPFL), CH-1015 Lausanne, Switzerland}
\affiliation{Center for Quantum Science and Engineering, EPFL, Lausanne, Switzerland}

\author{\\Riccardo Rende \orcidlink{0000-0001-5656-4241}}
\email{rrende@flatironinstitute.org}
\affiliation{Center for Computational Quantum Physics, Flatiron Institute, 162 5th Avenue, New York, NY 10010}

\author{Federico Becca \orcidlink{0000-0002-6142-2050}}
\affiliation{Dipartimento di Fisica, Universit\`a di Trieste, Strada Costiera 11, I-34151 Trieste, Italy}

\author{Giuseppe Carleo \orcidlink{0000-0002-8887-4356}}
\affiliation{Institute of Physics, \'{E}cole Polytechnique F\'{e}d\'{e}rale de Lausanne (EPFL), CH-1015 Lausanne, Switzerland}
\affiliation{Center for Quantum Science and Engineering, EPFL, Lausanne, Switzerland}

\date{\today}

\begin{abstract}
Materials design is an inverse problem: finding the couplings of a microscopic Hamiltonian whose ground state exhibits a desired property.
For strongly correlated quantum systems, this remains largely out of reach, as each candidate Hamiltonian requires its own costly simulation.
Here we show that Foundation Neural-Network Quantum States, single wave functions that encode the ground states of an entire family of Hamiltonians, make this search efficient.
Because the Hamiltonian couplings enter as neural-network inputs, any ground-state property can be optimized by gradient descent directly in coupling space.
We use this approach to search for nonmagnetic phases in two-dimensional spin models in coupling spaces of increasing dimensionality. Specifically, we focus on quantum spin liquids, exotic phases of matter in which quantum fluctuations prevent magnetic order at zero temperature.
In the largest search space considered, spanned by eight superexchange couplings and containing many paradigmatic frustrated spin systems, the optimization converges to the $J_1$-$J_2$-$\delta$ Heisenberg model when starting from a magnetic phase. We show that the ground state at the optimum set of couplings has neither magnetic nor valence-bond order in the thermodynamic limit, leading to a new quantum spin-liquid candidate. The approach is completely general and can be extended to other models and physical properties to optimize, demonstrating that 
Neural-Network Quantum States are tools not only for simulating quantum matter but also for designing it.
\end{abstract}

\maketitle

\section{Introduction}

A fundamental goal of materials science is to deliver materials with desired properties~\cite{marzari2021electronic,curtarolo2013high,jain2013commentary,basov2017towards,butler2018machine}.
This goal can be naturally framed as an \emph{inverse problem}~\cite{zunger2018inverse}: rather than computing the properties of a given system, one searches for the systems that realize prescribed ones.
In chemistry and materials science, this paradigm has mostly been pursued at the level of atomic structures and compositions, increasingly with machine-learning generative models for molecules~\cite{sanchez2018inverse} and crystalline solids~\cite{noh2019inverse,zeni2025generative}.
For strongly correlated systems, the natural formulation is instead at the level of the physical model: finding the couplings of the Hamiltonian whose ground state best exhibits the target behavior.
Previous approaches to this inverse model design have been limited to classical systems~\cite{okigami2024exploring}, to non-interacting models~\cite{inui2023inverse}, or to small interacting quantum systems accessible by exact diagonalization, where the goal is either to reconstruct a parent Hamiltonian from a given target state~\cite{chertkov2018computational,dupont2019eigenstate,fujita2018construction} or to optimize target properties such as entanglement~\cite{inui2024inverse} or topological order~\cite{fonseca2025gradient}.
An alternative route based on experimental quantum simulation has also been explored~\cite{kokail2026inverse}.
What remains missing is a scalable framework for \emph{ab initio} inverse model design in strongly interacting quantum systems targeting general physical properties.

Standard numerical methods for quantum many-body systems solve the \emph{forward} problem.
A Hamiltonian $\hat H$ is chosen based on phenomenological or first-principles calculations, its ground state is computed, and the resulting observables are compared with experiments.
Quantum Monte Carlo~\cite{ceperley1986quantum,foulkes2001quantum}, Density-Matrix Renormalization Group~\cite{white1992density}, Tensor Networks~\cite{verstraete2008matrix,orus2014practical}, and more recently Neural-Network Quantum States (NQS)~\cite{carleo2017solving} provide powerful tools for this task, each within its respective domain of applicability.
All of them, however, typically address a single Hamiltonian at a time.
A brute-force solution of the inverse problem would thus require solving the ground-state problem at many candidate points in coupling space and selecting the best.
While the cost of these repeated simulations can be reduced by reusing information from earlier computations~\cite{rath2025interpolating}, the size of the search still grows rapidly with the number of free couplings.
Moreover, isolated forward calculations do not, by themselves, provide a strategy for choosing the next candidate coupling.

\begin{figure*}
\centering
\includegraphics[width=1.0\linewidth]{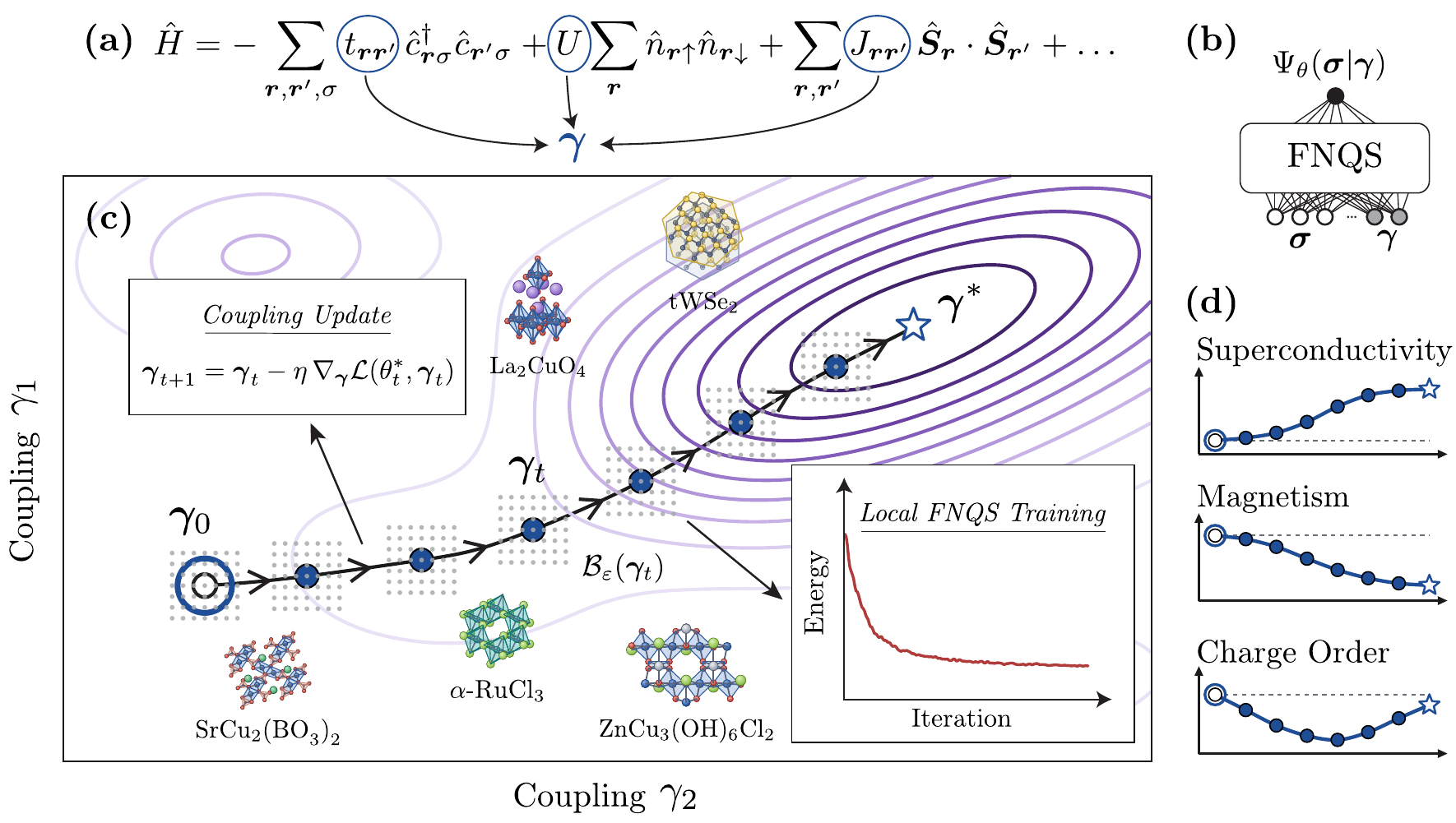}
\caption{\textbf{FNQS Inverse Model Design}.
Schematic of inverse model design based on FNQS.
\textbf{(a)} Example of a family of low-energy effective Hamiltonians, parametrized by the couplings $\boldsymbol{\gamma}$.
\textbf{(b)} Illustration of the FNQS ansatz, which takes both the many-body configuration $\boldsymbol{\sigma}$ and the couplings $\boldsymbol{\gamma}$ as inputs.
\textbf{(c)} Inverse design in coupling space.
Contour lines show the target property $\mathcal{L}$, from lighter (higher) to darker (lower) values, and material icons illustrate that different compounds correspond to different points of this space.
The trajectory shows the evolution of the couplings from an initial guess
$\boldsymbol{\gamma}_{0}$ (open circle) to the optimal model $\boldsymbol{\gamma}^{*}$ (star). 
Each iteration consists of the two stages described in the main text. \emph{(1) Local FNQS Training.} At the current iterate $\J_t$, the variational wave function $\Psi_\theta(\boldsymbol{\sigma}|\J)$ is optimized by VMC to represent the ground states of the Hamiltonians within the hypercube $\mathcal{B}_\varepsilon(\J_t)$ (grey points along the trajectory), by minimizing the average energy of~\cref{eq:fnqs} until convergence. \emph{(2) Coupling Update.} The gradient $\nabla_{\boldsymbol{\gamma}}\mathcal{L}(\theta^*,\boldsymbol{\gamma})$ is evaluated at the optimized parameters $\theta^*$ by differentiating the wave function with respect to the couplings, and the couplings are updated along the descent direction [see \cref{eq:jupdate}].
\textbf{(d)} Evolution of representative ground-state properties along the trajectory; depending on the choice of $\mathcal{L}$, each property can be enhanced or suppressed.
The figure is schematic and intended for illustration only; it does not represent quantitative results.
}
\label{fig:artistic_figure}
\end{figure*}

In this work, we introduce an efficient framework for \emph{ab initio} inverse model design based on Foundation Neural-Network Quantum States (FNQS)~\cite{rende2025,viteritti2025quantum,nazaryan2026qernel,foster2026ab,scherbela2024towards,gao2024neural,gerard2025transferable}, variational wave functions that represent the ground states of an entire family of Hamiltonians at once, with the couplings entering as explicit arguments alongside the many-body configuration.
Given a target family of Hamiltonians, the method searches for the member whose ground state optimizes a chosen physical property.
Because the couplings are inputs of the wave function, any property of the ground state becomes a differentiable function of them, and its gradient is obtained exactly, at the same cost as the variational forces of a standard ground-state calculation.
The family of Hamiltonians thus becomes a search space in which the optimal model is reached by gradient descent, without solving the forward problem on a grid of candidate couplings.
The approach is completely general, in that it applies to spin, bosonic, or fermionic systems and to any property computable on the variational state, and \emph{ab initio}, in the sense that it operates directly on the full many-body Hamiltonians defining the family.
To the extent that the couplings of the optimal Hamiltonian can be realized by tuning experimentally accessible control parameters, the procedure provides a route to designing materials with targeted properties from first principles.

We apply the method to the inverse design of nonmagnetic phases in two-dimensional quantum spin systems, with the aim of identifying candidate quantum spin liquid (QSL) states.
QSLs are magnetically disordered phases that preserve all the symmetries of the Hamiltonian and evade characterization by any local order parameter, sustaining instead long-range entanglement and fractionalized excitations~\cite{balents2010spin,savary2017quantum,zhou2017quantum}.
They are of central theoretical interest, having been linked to the mechanism of high-temperature superconductivity~\cite{anderson1987resonating} and proposed as a platform for topological quantum computation~\cite{kitaev2003fault}.
They are also actively sought in candidate materials, including herbertsmithite, organic salts on triangular lattices, and $\alpha$-RuCl$_3$~\cite{broholm2020quantum,norman2016colloquium}.
Yet the spin Hamiltonians that host them are known only in a handful of cases~\cite{kitaev2006, savary2017quantum}, typically identified by forward calculations at selected points in one- or two-dimensional coupling spaces~\cite{liu2022emergence, hu2013direct, iqbal2013gapless, iqbal2016, viteritti2023transformer_2d, chen2024empowering}.
Here, we search directly for such Hamiltonians by minimizing the magnetic order of the
ground state in progressively higher-dimensional spaces.
The largest space of eight superexchange couplings spans a broad family of Heisenberg models on the square lattice, including several examples of frustrated magnets~\cite{nomura2021dirac, shastry1981exact, canals2001square}.
Within this family, the optimization converges to the $J_1$-$J_2$-$\delta$ Heisenberg model, recently introduced in the context of altermagnetism~\cite{vsmejkal2022beyond,brekke2023two}.
At the selected couplings, previous studies predicted a nonmagnetic phase, but its nature had remained open~\cite{cichutek2025quantum,paul2026j_,djuric2026altermagnetism}.
We find that this state displays neither conventional magnetic nor valence-bond order, identifying it as a new QSL candidate.
The form of this Hamiltonian is determined solely by the optimization rather than assumed in advance, showing that inverse design can guide the discovery of quantum phases of matter.

\section{Methods}
\subsection{Variational Approach to Inverse Design}

Consider a many-body Hilbert space with configurations $\boldsymbol{\sigma}=(\sigma_1, \dots, \sigma_N)$ of $N$ discrete degrees of freedom, which may be spins, bosonic, fermionic occupation numbers, or similar, and a family of systems represented by Hamiltonians $\hat H(\J)$ labeled by a set of couplings $\J$.
An FNQS~\cite{rende2025,viteritti2025quantum} is a variational wave function $\Psi_\theta(\boldsymbol{\sigma}|\J)$, parametrized by a neural network with parameters $\theta$, that depends jointly on the many-body configuration and on the couplings, and that is optimized to approximate the ground state of $\hat H(\J)$ for many realizations of $\J$ simultaneously.
As in Variational Monte Carlo (VMC)~\cite{mcmillan1965ground,becca2017}, the optimal parameters are obtained by minimizing the energy, now averaged over a representative distribution $p(\J)$ of the couplings,
\begin{equation}
\label{eq:fnqs}
\theta^\ast = \arg\min_\theta\, \mathbb{E}_{\J\sim p(\J)}\!\left[
\frac{\bra{\Psi_\theta(\J)}\hat H(\J)\ket{\Psi_\theta(\J)}}
{\braket{\Psi_\theta(\J)}{\Psi_\theta(\J)}}\right].
\end{equation}

In practice, the previous expression and all other analogous expectation values, which we denote in the following by $\langle \ldots \rangle_{\J}$, are estimated by Monte Carlo sampling from $|\Psi_\theta(\boldsymbol{\sigma}|\J)|^2$~\cite{becca2017}.
For further details on this generalized energy-minimization framework, see Ref.~\cite{rende2025}.

Once the wave function represents the ground state across a representative set of Hamiltonians, the family itself becomes a search space: one can optimize a target property $\mathcal{L}$ over the couplings and thereby single out the Hamiltonian whose ground state is optimal with respect to it.
The property $\mathcal{L}$ may be any quantity computable on the variational state, for instance the magnetic order of a quantum spin model, the superconducting pairing of lattice fermions, or the pressure of electrons in continuous space.
The optimal system is then obtained by solving
\begin{equation}
\label{eq:inverse}
\J^{*} = \arg\min_{\J} \mathcal{L}(\theta^*, \J),
\end{equation}
where $\mathcal{L}(\theta^*, \J)$ denotes the property $\mathcal{L}$ evaluated on the optimized FNQS wave function ${\Psi_{\theta^*}(\boldsymbol{\sigma}|\J)}$; maximization is handled by minimizing $-\mathcal{L}$.
\cref{eq:inverse} admits many solution strategies, but to be effective the search should only visit couplings at which the variational state is an accurate ground state.

Our procedure starts from an initial guess $\J_{0}$ and iteratively updates the couplings to minimize $\mathcal{L}$, using two ingredients.
The first is the local training of the FNQS.
Let $\mathcal{B}_\varepsilon(\J)$ be the hypercube of side $\varepsilon$ centered on $\J$.
Training the FNQS on $\mathcal{B}_\varepsilon(\J)$ means minimizing the average energy in~\cref{eq:fnqs} by VMC, with the couplings drawn from a distribution $p(\J)$ with support on this hypercube.
The second ingredient is the gradient $\nabla_{\J}\mathcal{L}$ of the target property with respect to the couplings.

The iteration $t$ of the coupling optimization procedure then consists of two stages (see~\cref{fig:artistic_figure}):
\begin{enumerate}
    \item \emph{Local FNQS Training.} Train the FNQS by VMC on $\mathcal{B}_\varepsilon(\J_t)$ until convergence, obtaining the optimized parameters $\theta^{*}_t$.
    \item \emph{Coupling Update.} Estimate $\nabla_{\J}\mathcal{L}(\theta^{*}_t,\J_t)$ and update the couplings to decrease $\mathcal{L}$, using a gradient-based optimizer; for example, considering stochastic gradient descent (SGD) with learning rate $\eta$,
    \begin{equation}
    \label{eq:jupdate}
        \J_{t+1} = \J_t - \eta\, \nabla_{\J}\mathcal{L}(\theta^{*}_t,\J_t).
    \end{equation}
\end{enumerate}
The two stages are repeated until $\mathcal{L}$ converges, or until another stopping criterion is met, and the final couplings provide the estimate of $\J^{*}$ in~\cref{eq:inverse}.

We remark that the convergence of the coupling optimization can be further improved beyond SGD by performing Stochastic Reconfiguration (SR)~\cite{sorella1998green,
sorella2005wave} directly in coupling space.
This amounts to replacing the gradient in the update rule of~\cref{eq:jupdate} with the corresponding natural gradient~\cite{Amari1998,Amari2018}, obtained by preconditioning $\nabla_{\J} \mathcal{L}$ with the inverse of the real part of the quantum geometric tensor,
\begin{equation}
\label{eq:qgt_couplings}
    \mathcal{S}_{kl}(\theta, \J) = \mathrm{Re}\Big[\langle \overline{\hat{O}^{\dagger}_{\gamma_k}} \; \overline{\hat{O}_{\gamma_l}} \rangle_{\J} \Big] ,
\end{equation}
evaluated at the same parameters and couplings as the gradient.
In the previous expression, $\hat{O}_{\gamma_k}$ is the diagonal operator with entries $O_{\gamma_k}(\boldsymbol{\sigma}|\J) = \partial_{\gamma_k} \log \Psi_{\theta}(\boldsymbol{\sigma}|\J)$. 
In this case the derivative is taken with respect to the couplings which are an \emph{input} of the neural network rather than one of its variational parameters, and is likewise obtained by backpropagation~\cite{rende2025}.
Here and in what follows, a bar over an operator denotes its connected version, obtained by subtracting its expectation value.
The SR update is expected to converge faster than SGD and Adam~\cite{kingma2017adam}, particularly in its momentum-based variants~\cite{goldshlager2024kaczmarz,gu2026solving}. In this work we employ one such variant, the MARCH optimizer~\cite{gu2026solving}, which we found to converge more reliably than the other schemes we tested.
The momentum accumulated by MARCH also helps the optimization jump over phase transition points, which can be pathological for the natural gradient, since $\mathcal{S}$ coincides with the generalized fidelity susceptibility~\cite{wang2015fidelity,rende2025} that diverges there in the thermodynamic limit~\cite{zanardi2007,campos2007}, suppressing the update step.

An essential feature of the inverse design procedure is that the FNQS is trained only locally, rather than globally over the whole coupling space.
Since each coupling update depends only on evaluations at the current couplings $\J_t$, the FNQS must represent the ground states only in their neighborhood $\mathcal{B}_\varepsilon(\J_t)$, and the training region can follow the optimization trajectory.
This restriction allows the energy to be minimized with much higher accuracy, yielding reliable estimates of $\mathcal{L}$ and $\nabla_{\J}\mathcal{L}$.
The side $\varepsilon$ cannot be taken arbitrarily small, however: the hypercube must be wide enough for the wave function to resolve how the ground state varies with $\J$, which is necessary to compute $\nabla_{\J}\mathcal{L}$.
Typically, $\varepsilon$ can be chosen to be of the order of the coupling displacement per update.
Because the couplings are arguments of the FNQS, the gradient $\nabla_{\J}\mathcal{L}$ is obtained by differentiating the network with respect to $\J$, at the cost of a single backward pass.
Moreover, since consecutive hypercubes are close in coupling space, initializing each VMC training with the parameters of the previous iteration substantially reduces its cost compared with training from scratch.

The framework we have introduced is general and applies to any system for which such a family of Hamiltonians can be defined. 
In particular, it can be employed within a materials design workflow, in combination with \emph{ab initio} downfolding
techniques~\cite{imada2010electronic,PhysRevB.70.195104}, which derive effective
Hamiltonians describing the physics of real materials at low energy [see~\cref{fig:artistic_figure}(a)].
The procedure would be to start from a family of materials, use downfolding to fix the form of the effective Hamiltonian describing them, treat its couplings as free variables, and use FNQS to optimize the target property over the resulting coupling space.
If the optimized couplings can be realized experimentally by tuning accessible control parameters, the resulting Hamiltonian identifies a candidate material with the desired property.

\subsection{Inverse Design of Nonmagnetic Phases}
\label{sec:nonmagnetic_phases}
As an application of the method, we consider quantum spin-$\tfrac{1}{2}$ systems on a two-dimensional lattice of $N$ sites.
Our goal is to identify candidates for QSL states, highly entangled phases in which strong quantum fluctuations and frustration prevent magnetic order from developing down to zero temperature~\cite{balents2010spin, savary2017quantum}.
We therefore look for Hamiltonians whose ground states have minimal magnetic correlations.
For simplicity, we assume translational invariance, but the generalization to systems without this symmetry is straightforward.
To identify nonmagnetic phases, a suitable choice for $\mathcal{L}$ is the Edwards-Anderson order parameter~\cite{parisi1983, baity2015inherent, ren2023, viteritti2025quantum},
\begin{equation}
\mathcal{L}(\theta, \J) = \frac{1}{N} \sum_{\boldsymbol{r}}
\langle \hat{\bm{S}}_{\boldsymbol{0}} \cdot
\hat{\bm{S}}_{\boldsymbol{r}}\rangle_{\J}^2 \equiv Q^2 ,
\label{eq:qea}
\end{equation}
where $\hat{\bm{S}}_{\bm r} = (\hat{S}^x_{\boldsymbol{r}}, \hat{S}^y_{\boldsymbol{r}},
\hat{S}^z_{\boldsymbol{r}})$ is the spin operator at site $\boldsymbol{r}$, and the sum runs over all lattice sites.
This choice is justified by the identity ${Q^2 = \sum_{\boldsymbol{k}} S(\boldsymbol{k}) S(-\boldsymbol{k})/N^2}$,
where $S(\boldsymbol{k})$ is the static spin structure factor:
\begin{equation}
\label{eq:structure_factor}
S(\boldsymbol{k}) = \sum_{\boldsymbol{r}} e^{i \boldsymbol{k} \cdot \boldsymbol{r}} \langle \hat{\bm{S}}_{\boldsymbol{0}} \cdot
\hat{\bm{S}}_{\boldsymbol{r}}\rangle_{\J}.
\end{equation}

Magnetic order shows up as a Bragg peak in $S(\boldsymbol{k})$, whose weight grows proportionally to $N$ at the ordering wave vector; since $Q^2$ weighs every wave vector equally, minimizing it suppresses all such peaks at once, without presupposing where in the Brillouin zone they might occur.
This is essential when the ordering pattern is not known in advance, as happens whenever several magnetic orders compete.
Moreover, being the spin-glass order parameter, $Q^2$ is sensitive to patterns that carry no Bragg peak at all.

A magnetic state has $Q^2$ approaching a nonzero value in the thermodynamic limit, whereas $Q^2 \to 0$ for $N \to \infty$ signals the absence of magnetic order and thus identifies a candidate QSL.
The criterion is necessary but not sufficient: a valence-bond solid (VBS) state also has vanishing $Q^2$.
Once the minimum of~\cref{eq:qea} has been reached, a VBS can be ruled out by inspecting the connected dimer-dimer correlations, which decay to zero at large separation in a QSL but saturate to a finite value in a VBS.
Alternatively, a dimer order parameter could be included directly in the target property of~\cref{eq:qea}, penalizing both types of order at once; this would however require the estimation of four-body correlators at every iteration,
raising the computational cost appreciably.

The gradient of~\cref{eq:qea} with respect to $\J$ follows from the chain rule together with the standard VMC expression for the derivative of an expectation value, which takes the form of a covariance between the observable and the logarithmic derivative of the ansatz:
\begin{equation}
    \partial_{\gamma_k} \mathcal{L}(\theta, \J) = \frac{4}{N} \sum_{\boldsymbol{r}} \big\langle \hat{\bm{S}}_{\boldsymbol{0}} \cdot \hat{\bm{S}}_{\boldsymbol{r}} \big\rangle_{\J} \; \mathrm{Re} \bigg[\big\langle \overline{\hat{O}^{\dagger}_{\gamma_k}} \; \overline{\hat{\bm{S}}_{\boldsymbol{0}} \cdot \hat{\bm{S}}_{\boldsymbol{r}}} \big\rangle_{\J}\bigg].
    \label{eq:qeagrad}
\end{equation}

\section{Results}
The families of Hamiltonians we explore are generalizations of the spin-$\tfrac{1}{2}$ Heisenberg model on the two-dimensional square lattice, with nearest-neighbor coupling $J_1$ and next-nearest-neighbor couplings along the diagonals.
We denote by $L$ the linear size of the cluster, for a total of $N = L^2$ sites.
In all the calculations, we use periodic boundary conditions and restrict to antiferromagnetic interactions.
Throughout, we progressively enlarge the freedom allowed in the diagonal couplings, which thus constitute the free
variables $\boldsymbol{\gamma}$ of the inverse design.
We begin, as a proof of concept, with the standard case of a single diagonal coupling (see \cref{sec:one_coup}); we then allow the two diagonals of each plaquette to differ, giving two independent couplings (see \cref{sec:two_coup}); finally, we consider eight independent diagonal couplings within a $2\times2$ unit cell (see \cref{sec:eight_coup}).
Each enlargement widens the space of models available to the search.
For the one- and two-coupling cases, we take $p(\J)$ to be a discrete distribution $p(\J) = (1 / \mathcal{R} )\sum_{k=1}^{\mathcal{R}} \delta(\J - \J_k)$, where the $\mathcal{R}$ couplings $\J_k$ lie on a regular grid inside the hypercube.
For the eight-coupling case, we instead take $p(\J)$ to be uniform over the hypercube, since covering it with a grid would require a number of models growing exponentially with the number of couplings.

We adopt the FNQS architecture of Ref.~\cite{rende2025}, which builds on the transformer previously applied to spin models~\cite{viteritti2023transformer_1d, viteritti2023transformer_2d}. The network has $n_l=6$ layers, $h=12$ attention heads, and embedding dimension $d=72$, and its attention mechanism is modified with a spatial bias~\cite{viteritti2026approaching}. Moreover, to better capture the different sign structures of the ground states explored during the inverse design, we supplement the architecture with a coupling-dependent reference sign structure derived from the solution of the corresponding classical spin model.
Details on this construction are provided in Appendix~\ref{sec:sign}.

\subsection{One Coupling: $J_1$-$J_2$ Heisenberg Model}\label{sec:one_coup}
The first Hamiltonian we consider is the $J_1$-$J_2$ Heisenberg model on the square lattice,
\begin{equation}
\label{eq:j1j2_model}
\hat H = J_1 \sum_{\langle \boldsymbol{r}, \boldsymbol{r^\prime}\rangle} \Sv_{\boldsymbol{r}} \cdot \Sv_{\boldsymbol{r}^\prime} + J_2 \sum_{\langle\langle \boldsymbol{r}, \boldsymbol{r^\prime}\rangle\rangle} \Sv_{\boldsymbol{r}}\cdot\Sv_{\boldsymbol{r}^\prime} ,
\end{equation}
where the two sums run over nearest- and next-nearest-neighbor
pairs, respectively, and the only free coupling is the frustrating ratio $J_2/J_1$.
The phase diagram hosts two well-established ordered phases: N\'eel antiferromagnetic
order for $J_2/J_1 \lesssim 0.4$ and collinear stripe order for $J_2/J_1 \gtrsim 0.6$.
Between them lies a nonmagnetic regime, $0.4 \lesssim J_2/J_1 \lesssim 0.6$, whose nature has long been debated~\cite{gong2014plaquette,hu2013direct,choo2019two,capriotti,capriotti2}.
More recent studies suggest that this regime comprises a QSL followed by a VBS~\cite{liu2022gapless,nomura2021dirac,sandvik,ferrari}.

\begin{figure}
\centering
\includegraphics[width=\linewidth]{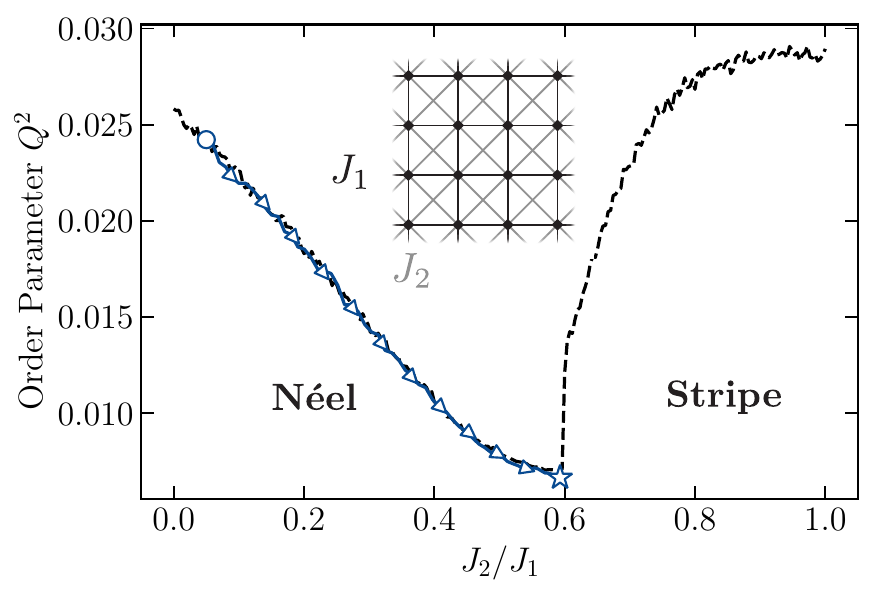}
\caption{
\textbf{$\boldsymbol{J_1}$-$\boldsymbol{J_2}$ Heisenberg Model}.
Inverse design of a nonmagnetic phase in the frustrated $J_1$-$J_2$ Heisenberg model of~\cref{eq:j1j2_model} with a single free coupling $J_2/J_1$, on a $12 \times 12$ lattice. 
The dashed black line shows the Edwards-Anderson order parameter $Q^2$, evaluated over the whole range $0 \leq J_2/J_1 \leq 1$ by a global FNQS calculation, which provides the reference landscape.
The Néel and stripe phases are explicitly indicated.
The blue trajectory shows the coupling optimization, started from the N\'eel phase at $J_2/J_1 = 0.05$ (open circle) and converging to the minimum of $Q^2$ at $J_2/J_1 \approx 0.6$ (star), within the nonmagnetic window where the QSL is expected; the arrowheads indicate the direction of the minimization.
Inset: lattice geometry, with nearest-neighbor bonds $J_1$ (black) and next-nearest-neighbor bonds $J_2$ (grey).}
\label{fig:fig1}
\end{figure}

This one-coupling case serves as a benchmark: it tests whether the minimization of $Q^2$ through the FNQS locates the nonmagnetic window without any prior knowledge of the phase diagram, before the method is applied to coupling spaces where no reference is available. \cref{fig:fig1} shows the optimization trajectory on a $12 \times 12$ cluster, started from a point deep in the N\'eel phase, superimposed on the $Q^2$ landscape obtained from a standard global FNQS calculation over the whole range $0 \leq J_2/J_1 \leq 1$.
The order parameter descends monotonically and converges to a minimum at $J_2/J_1\approx0.6$, inside the nonmagnetic window, confirming that the inverse-design procedure recovers the known answer.

\begin{figure*}[htbp]
\centering
\includegraphics[width=\linewidth]{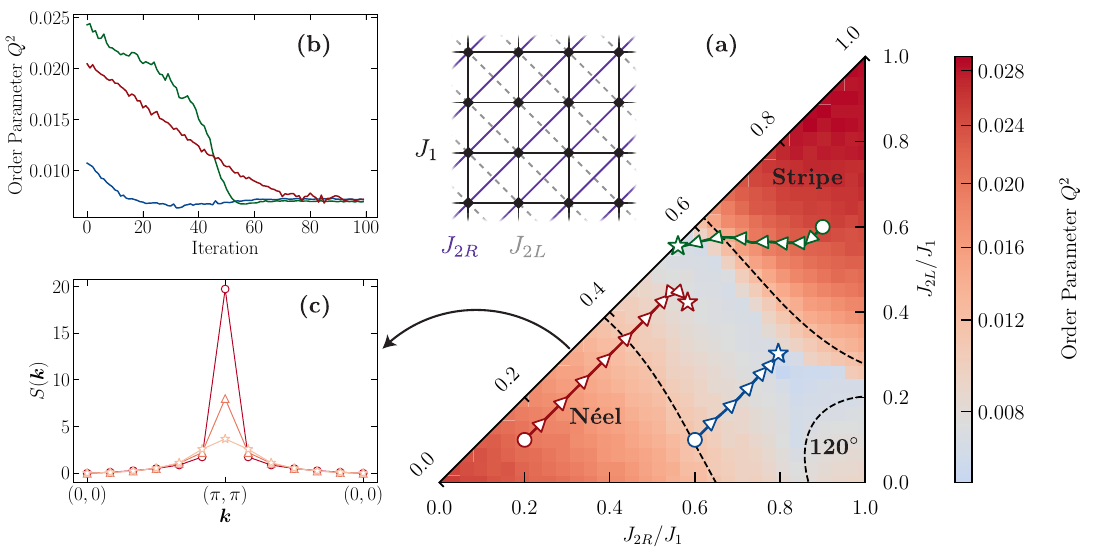}
\caption{\textbf{Generalized $\boldsymbol{J_1}$-$\boldsymbol{J_2}$ Heisenberg Model}.
Inverse design of nonmagnetic phases in the generalized frustrated Heisenberg model of~\cref{eq:generalized_j1j2}, with two independent next-nearest-neighbor couplings $J_{2R}/J_1$ and $J_{2L}/J_1$, on a $12 \times 12$ lattice. 
\textbf{(a)} Edwards-Anderson order parameter $Q^2$, evaluated over the region $0 \leq J_{2L}/J_1 \leq J_{2R}/J_1 \leq 1$ by a global FNQS calculation, which provides the reference landscape; the complementary half of the phase diagram follows from the symmetry $J_{2R} \leftrightarrow J_{2L}$.
Dashed lines give an indicative estimate of the extent of the N\'eel, stripe, and $120^\circ$ ordered phases, identified from the peaks of the structure factor at $\boldsymbol{k}=(\pi,\pi)$, $(\pi,0)$ and $(0,\pi)$, and $\pm(2\pi/3,2\pi/3)$, respectively; each boundary is drawn where the corresponding order parameter falls below $50\%$ of its maximum value over the phase diagram.
The three trajectories show independent coupling optimizations started from three different initial points, each marked by an open circle at its start and a star at its endpoint; the arrowheads indicate the direction of the minimization.
\textbf{(b)} Decrease of $Q^2$ along the three trajectories as a function of the iteration of the coupling update.
\textbf{(c)} Static spin structure factor $S(\boldsymbol{k})$ along the diagonal of the Brillouin zone at wave vectors $\boldsymbol{k} = (n \frac{2\pi}{L}, n \frac{2\pi}{L})$ with $n = 0, \ldots, L$, computed at three stages of the red trajectory: the initial point (circles), an intermediate point (triangles), and the final point (stars).
Inset: lattice geometry, with nearest-neighbor bonds $J_1$ (black) and the two inequivalent next-nearest-neighbor bonds $J_{2R}$ (purple) and $J_{2L}$ (grey, dashed).
}
\label{fig:fig2}
\end{figure*}

\subsection{Two Couplings: Generalized $J_1$-$J_2$ Heisenberg Model}\label{sec:two_coup}
We next consider a generalization of~\cref{eq:j1j2_model} in which the two diagonals of each plaquette carry different next-nearest-neighbor couplings:
\begin{equation}
\begin{split}
\label{eq:generalized_j1j2}
\hat H &= J_1 \sum_{\langle \boldsymbol{r}, \boldsymbol{r^\prime} \rangle} \Sv_{\boldsymbol{r}}\cdot\Sv_{\boldsymbol{r}^\prime} + J_{2R}\sum_{\boldsymbol{r}}\Sv_{\boldsymbol{r}}\cdot\Sv_{\boldsymbol{r} + \hat x + \hat y} \\ 
&+ J_{2L}\sum_{\boldsymbol{r}} \Sv_{\boldsymbol{r}}\cdot\Sv_{\boldsymbol{r} + \hat x - \hat y},
\end{split}
\end{equation}
where $\hat x$ and $\hat y$ denote the two perpendicular directions of the square lattice.
This model was proposed in Ref.~\cite{rende2025} to investigate out-of-distribution generalization in FNQS.
In this case, we have two free couplings, $J_{2R}/J_1$ and $J_{2L}/J_1$.
A reflection of the lattice about the $x$ (or $y$) axis exchanges the two diagonals, mapping~\cref{eq:generalized_j1j2} onto the same Hamiltonian with $J_{2R}$ and $J_{2L}$ interchanged.
The two are unitarily equivalent, so it suffices to explore the region $J_{2L} \leq J_{2R}$, the complementary region following by reflection.

This family of systems interpolates between two well-known models.
For $J_{2R} = J_{2L}$ it reduces to the standard $J_1$-$J_2$ Heisenberg model of~\cref{eq:j1j2_model}, whereas for $J_{2L} = 0$ the remaining diagonal turns the square lattice into an anisotropic triangular lattice, whose Heisenberg model is likewise expected to host a nonmagnetic phase around $J_{2R}/J_1 \approx0.8$~\cite{weng_triangular,ghorbani_triangular,juraj_triangular}.
It is therefore natural to ask whether the nonmagnetic windows of the square $J_1$-$J_2$ and anisotropic triangular models form two disconnected pockets or two limits of a single region, and, more generally, what happens in between, for $J_{2L} \neq J_{2R}$ with both couplings nonzero.

To address this, we first performed a standard global FNQS calculation over the region $0 \leq J_{2L}/J_1 \leq J_{2R}/J_1 \leq 1$.
As shown in~\cref{fig:fig2}(a), the phase diagram contains three magnetically ordered phases, displaying N\'eel, stripe, and $120^\circ$ order.
These are identified from the corresponding order parameters, obtained from the peak of the structure factor $S(\boldsymbol{k})$ of~\cref{eq:structure_factor} at $\boldsymbol{k} = (\pi, \pi)$ for the N\'eel order,
at $\boldsymbol{k} = (\pi, 0)$ and $(0, \pi)$ for the stripe phase, and at $\boldsymbol{k} = \pm(2\pi/3, 2\pi/3)$ for the $120^\circ$ order.
The three ordered regions are separated by a wide
region of suppressed $Q^2$, which seems to connect the QSL of the $J_1$-$J_2$ model at $J_2/J_1 \approx 0.6$ to that of the anisotropic triangular limit at $J_{2L}/J_1 = 0$ and $J_{2R}/J_1 \approx 0.8$.

We then performed three independent inverse-design optimizations, started from different points of the phase diagram.
As shown in~\cref{fig:fig2}(a), all of them converge to models lying in the nonmagnetic region, but to distinct points within it: a direct signature of the near-degeneracy of $Q^2$ along the region, and an illustration of the fact that, when the minimum is not isolated, the inverse design returns one representative of the optimal set of models rather than a unique solution.
Consistently, panel (b) shows that $Q^2$ decreases along all three trajectories and settles at comparable plateau values.
Panel (c) illustrates how this suppression of the magnetic correlations comes about for the trajectory starting in the N\'eel phase, displaying the structure factor $S(\boldsymbol{k})$ along the diagonal of the Brillouin zone at three stages of the inverse-design optimization.
The Bragg peak at $\boldsymbol{k}=(\pi,\pi)$ is progressively suppressed as the couplings are driven towards the nonmagnetic region, signaling the gradual disappearance of the antiferromagnetic order.

\begin{figure*}[htbp]
\centering
\includegraphics[width=\linewidth]{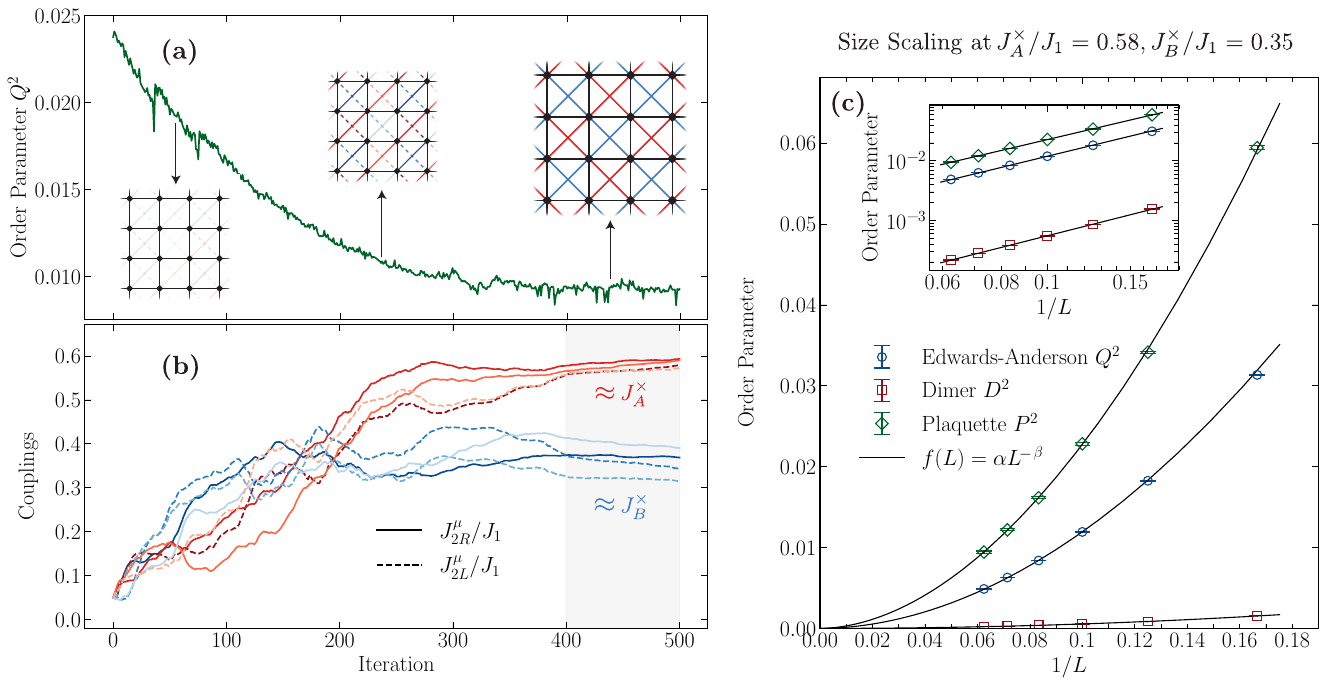}
\caption{$\boldsymbol{J_1}$-$\boldsymbol{J_2}$ \textbf{Heisenberg Model with $\boldsymbol{2\times2}$ Unit Cell}.
Inverse design of nonmagnetic phases in the frustrated Heisenberg model of~\cref{eq:H3} with eight independent next-nearest-neighbor couplings $\{J_{2R}^{\mu}, J_{2L}^{\mu}\}_{\mu=1}^{4}$ within a $2\times2$ unit cell.
\textbf{(a)} Minimization of the Edwards-Anderson order parameter $Q^2$ on a $12 \times 12$ lattice, starting from all eight couplings equal to $J_{2R}^{\mu}/J_1 = J_{2L}^{\mu}/J_1 = 0.05$, deep in the N\'eel phase.
The snapshots show the model at three stages of the
optimization, with distinct colors identifying the independent couplings and color intensity encoding their magnitudes.
\textbf{(b)} Evolution of the eight couplings along the same optimization, with solid lines for $J_{2R}^{\mu}/J_1$ and dashed lines for $J_{2L}^{\mu}/J_1$.
After an initial transient, the couplings separate into two groups of four (red and blue), each approaching a common plateau in the shaded region.
At convergence, the model is thus approximately described by two effective couplings, $J_{A}^{\times}$ and $J_{B}^{\times}$, arranged in a checkerboard pattern.
\textbf{(c)} Finite-size scaling of the Edwards-Anderson order parameter $Q^2$, the dimer order parameter $D^2$, and the plaquette order parameter $P^2$, for the effective model at $J_{A}^{\times}/J_1 = 0.58$ and $J_{B}^{\times}/J_1 = 0.35$, the values obtained by averaging the couplings within each group at convergence.
Square lattices of size from $6 \times 6$ to $16 \times 16$ are considered.
Solid lines are power-law fits $f(L) = \alpha L^{-\beta}$; the inset shows the same data in logarithmic scale.}
\label{fig:fig3}
\end{figure*}

\subsection{Multiple Couplings: $J_1$-$J_2$ Heisenberg Model with $2 \times 2$ Unit Cell}\label{sec:eight_coup}
\label{sec:eight_coupling}
Finally, we scale the inverse design to a genuinely high-dimensional coupling space.
We divide the square lattice into $2 \times 2$ unit cells and let each diagonal bond within the cell vary independently, for a total of eight free couplings. 
Denoting by $\deltav_\mu \in \{(0,0), (1,0), (0,1), (1,1)\}$, with $\mu = 1, \ldots, 4$, the positions of the four sites within the unit cell, and by $\Rv = 2n_x \hat x + 2n_y \hat y$, with $n_x, n_y = 0, \ldots, L/2 - 1$, the positions of the $N/4$ unit cells, the Hamiltonian can be written as:
\begin{equation}
\begin{split}
\hat H &= J_1 \sum_{\langle \boldsymbol{r}, \boldsymbol{r^\prime} \rangle} \Sv_{\boldsymbol{r}}\cdot\Sv_{\boldsymbol{r}^\prime} + \sum_{\Rv, \mu}  \Big[\, J_{2R}^{\mu}\, \Sv_{\Rv+\deltav_\mu}\cdot\Sv_{\Rv+\deltav_\mu+\hat x+\hat y} \\
&+ J_{2L}^{\mu}\, \Sv_{\Rv+\deltav_\mu}\cdot\Sv_{\Rv+\deltav_\mu+\hat x-\hat y} \,\Big].
\label{eq:H3}
\end{split}
\end{equation}

This family of Hamiltonians, with free couplings $\{J_{2R}^\mu, J_{2L}^\mu\}_{\mu=1}^{4}$, contains several of the most studied frustrated spin models as special slices: besides the models of~\cref{sec:one_coup,sec:two_coup}, it includes the Shastry-Sutherland~\cite{shastry1981exact,PhysRevLett.112.147203,koga2000quantum,PhysRevB.87.115144,PhysRevX.9.041037,yang2022quantum,PhysRevB.105.L041115,PhysRevLett.133.026502,viteritti2023transformer_2d} and the checkerboard~\cite{PhysRevB.69.214427,PhysRevB.65.140407,PhysRevB.67.054411,chan2011tensor,starykh2005anisotropic,bishop2012frustrated,canals2001square,zou2024nearly} models, together with the continuum of interpolations between them.
An eight-dimensional coupling space lies far beyond the reach of a brute-force approach, which would require an independent ground-state calculation at each point of a grid whose size grows exponentially with the number of couplings.
It therefore provides a stringent test of the inverse-design procedure, and, more importantly, a search space large enough to contain nonmagnetic models beyond the idealized ones already known, including less symmetric Hamiltonians that may be closer to those realized in actual materials.

\Cref{fig:fig3}(a) shows the minimization of $Q^2$ in the eight-dimensional coupling space on a $12 \times 12$ cluster, starting from a configuration where all couplings take the same small value, namely deep in the
N\'eel phase close to the unfrustrated Heisenberg model.
Additionally, the snapshots of the model at three stages of the optimization are reported.
Panel (b) shows the corresponding evolution of the eight couplings.
We observe that all the couplings initially grow and, after passing through a transient in which they take distinct nonzero values, split into two groups of four, each converging around a plateau value.
This behavior suggests that the eight-coupling model design is spontaneously converging to an effective system that can be approximately described by only two couplings.
The two groups occupy alternating plaquettes in a checkerboard pattern, as shown by the last snapshot of the system in panel (a).
We denote by $J_{A}^{\times}$ and $J_{B}^{\times}$ the two effective couplings on the alternating plaquettes, and estimate them by averaging the couplings within each group at convergence, obtaining approximately $J_{A}^{\times}/J_1 = 0.58$ and $J_{B}^{\times}/J_1 = 0.35$.

The effective model identified by the eight-coupling inverse design interpolates between the standard $J_1$-$J_2$ Heisenberg model, recovered when $J_{A}^{\times} = J_{B}^{\times}$, and the checkerboard (or planar pyrochlore) model, obtained when either of the two couplings vanishes.
The latter has been studied extensively with both analytical and numerical techniques, which have revealed a competition between N\'eel order, stripe order, and several types of VBS, depending on the coupling on the crossed plaquettes~\cite{PhysRevB.69.214427,PhysRevB.65.140407,PhysRevB.67.054411,chan2011tensor,starykh2005anisotropic,bishop2012frustrated,canals2001square,zou2024nearly}.
The interpolating model itself, known in the literature as the $J_1$-$J_2$-$\delta$ Heisenberg model, has recently been proposed as a minimal model for altermagnetism~\cite{vsmejkal2022beyond,brekke2023two}, with possible realizations in iron oxychalcogenides~\cite{zhu2010band} and in ultracold fermionic atoms in optical lattices with strong on-site interactions~\cite{das2024realizing}. Spin-wave analyses~\cite{cichutek2025quantum,paul2026j_} predict an extended magnetically disordered region separating the altermagnetic phase from the stripe phase, while numerical studies~\cite{djuric2026altermagnetism,liu2025quantum} have characterized the altermagnetic region in detail and reported a bond-nematic phase with coexisting symmetry-protected topological order at a single point inside the disordered region.
The values of the couplings identified by the inverse design place the model inside this predicted magnetically disordered region. 
We emphasize that no constraint was imposed on the couplings during the optimization: the minimization of the magnetic correlations measured by $Q^2$ alone drove the system toward this structure, breaking the uniformity of the initial configuration. 
Remarkably, an unconstrained search in a high-dimensional coupling space thus singles out, without any prior input, a model with a symmetric and non-trivial spatial pattern. 
However, we cannot rule out the existence of other optimal models reached from different initializations in coupling space.

\subsection{Nature of the Optimal Ground State in the Multiple-Coupling Case}
Having identified a candidate model with a nonmagnetic ground state, we now characterize
the nature of this phase, and in particular distinguish between a QSL and a VBS.
To this end, we perform a size scaling of several different order parameters and examine their
behavior in the thermodynamic limit.
In practice, we carry out VMC optimizations of the effective model at fixed couplings
$J_{A}^{\times}/J_1 = 0.58$ and $J_{B}^{\times}/J_1 = 0.35$ on $L \times L$ clusters of linear size $L = 6$ to $L = 16$, and fit the resulting order parameters as a function of $1/L$ to
extrapolate for $L\rightarrow\infty$.

Besides the Edwards-Anderson parameter $Q^2$, we consider two order parameters to rule out possible valence-bond order.
The first is the analogue of $Q^2$ for dimer correlations~\cite{wu2024unveiling,wu2019randomness,nomura2021dirac},
\begin{equation}
\label{eq:dimer}
    D^2 = \frac{N}{N_{b}^2} \sum_{\boldsymbol{r}} \sum_{\boldsymbol{\delta}, \boldsymbol{\delta}^\prime} \langle \hat B_{\boldsymbol{0}}^{\boldsymbol{\delta}} \hat B_{\boldsymbol{r}}^{\boldsymbol{\delta}^\prime} \rangle^2,
\end{equation}
where $\hat B_{\boldsymbol{r}}^{\boldsymbol{\delta}} = \Sv_{\boldsymbol{r}} \cdot \Sv_{\boldsymbol{r} + \boldsymbol{\delta}} - \langle \Sv_{\boldsymbol{r}} \cdot \Sv_{\boldsymbol{r} + \boldsymbol{\delta}} \rangle$ is the fluctuation of the bond energy on the nearest-neighbor bond from $\boldsymbol{r}$ along $\boldsymbol{\delta}$ and $N_b$ is the total number of bonds.
We consider $\boldsymbol{\delta} \in \{\hat x, \hat y\}$, such that $N_b = 2N$.
$D^2$ remains finite in the thermodynamic limit whenever the bond energies
develop long-range order, as in a columnar VBS.

To address plaquette order as well, we consider the connected correlation function~\cite{viteritti2023transformer_2d,yang2022quantum}
\begin{equation}
\label{eq:corr_func_plaquette}
    C_{P}(\boldsymbol{r}) = \langle \hat{\mathcal{P}}(\boldsymbol{r})\, \hat{\mathcal{P}}(\boldsymbol{0}) \rangle - \langle \hat{\mathcal{P}}(\boldsymbol{r}) \rangle \langle \hat{\mathcal{P}}(\boldsymbol{0}) \rangle,
\end{equation}
where $\hat{\mathcal{P}}(\boldsymbol{r}) = [\hat P(\boldsymbol{r}) + \hat P^{-1}(\boldsymbol{r})]/2$ and $\hat{P}(\boldsymbol{r})$ cyclically permutes the four spins of the square plaquette whose lower-left site is $\boldsymbol{r}$.
Denoting by $C_{P}(\boldsymbol{k})$ the Fourier transform of~\cref{eq:corr_func_plaquette}, defined with the same conventions as in~\cref{eq:structure_factor}, plaquette order appears as peaks at $\boldsymbol{k} = (\pi, 0)$ and $(0, \pi)$. 
Therefore, we can define the plaquette order parameter as
\begin{equation}
\label{eq:plaquette}
    P^2 = \frac{1}{2}\big[\, C_P(\boldsymbol{k} = (\pi, 0)) +  C_P(\boldsymbol{k} = (0, \pi)) \,\big] .
\end{equation}

In~\cref{fig:fig3}(c) we report $Q^2$, $D^2$, and $P^2$ for the ground state of the effective model at different system sizes. 
We observe that all three decrease monotonically with system size and are well described by a power law in $1/L$, as confirmed by the linear behavior in the log-log inset, and thus extrapolate to zero in the thermodynamic limit.  
This calculation provides numerical evidence that the ground state of the nontrivial effective model found by our inverse design displays neither magnetic, nor dimer, nor plaquette order, strongly suggesting a QSL ground state.
Moreover, one can show in general that $Q^2 \leq S_{\max}/N$, where $S_{\max}$ is the peak of the structure factor.
Since we empirically find $Q^2 = \alpha L^{-\beta}$ with $\alpha > 0$ and $\beta < 2$, it follows that $S_{\max}$ diverges in the thermodynamic limit.
This provides first evidence that the QSL ground state of this model is gapless.

\section{Conclusions and Outlook}
In this work, we have introduced a general framework for \emph{ab initio} inverse model design based on Foundation Neural-Network Quantum States.
Because an FNQS takes the Hamiltonian couplings as inputs, any ground-state property becomes a differentiable function of them, and alternating energy minimizations in small regions of the coupling space with gradient-based coupling updates drives the search towards an optimal Hamiltonian, without solving the ground-state problem over a grid of candidate couplings.
The procedure is therefore well suited to high-dimensional coupling spaces, where brute-force exploration becomes unaffordable.

We have applied the framework to the inverse design of nonmagnetic phases in frustrated two-dimensional Heisenberg models with an increasing number of free couplings, aiming
to identify quantum spin liquid states.
In the largest coupling space, the optimization converges to the $J_1$-$J_2$-$\delta$ Heisenberg model, at couplings where earlier studies predicted a nonmagnetic phase of undetermined nature~\cite{cichutek2025quantum,paul2026j_,djuric2026altermagnetism}.
There, the magnetic, dimer, and plaquette order parameters of the ground state all extrapolate to zero in the thermodynamic limit, identifying this phase as a new QSL candidate.
Although this Hamiltonian had been proposed in the context of altermagnetism~\cite{vsmejkal2022beyond,brekke2023two}, its structure emerged here from an unconstrained search over a coupling space far too large for grid-based exploration, making the QSL candidate a prediction of the method itself.

Several directions for future developments naturally follow.
On the side of the target property, the loss function can be extended to penalize different types of order at once, for instance by including dimer correlations as discussed in connection with~\cref{eq:qea}, or to include additional terms that restrict the couplings to experimentally accessible ranges.
On the side of the Hamiltonian family, the search can be extended to further-neighbor and anisotropic interactions, such as Dzyaloshinskii-Moriya or Kitaev-type couplings, to other lattice geometries, such as the triangular, kagome, and honeycomb lattices, and to larger unit cells.
Beyond quantum magnetism, the framework applies without modification to fermionic and bosonic systems: a natural next target is the design of models with enhanced superconducting pairing correlations in Hubbard-like Hamiltonians, where NQS have recently reached state-of-the-art accuracy~\cite{Roth2025SCNQS,gu2026solving,viteritti2026beyond,rende2026superconductivity}, as well as properties of lattice bosons~\cite{denis2024accurate}, electrons in continuous space~\cite{ferminet,hermann2020deep,pescia,PhysRevLett.130.036401,PhysRevX.14.021030,qxc3-bkc7}, and excited-state properties~\cite{pfau2024accurate,entwistle2023electronic,hendry2025grassmannvariationalmontecarlo}.
Finally, connecting the optimal model couplings to experimentally controllable parameters is essential to turn inverse model design into materials design.
This can be achieved by combining the framework with first-principles derivations of effective Hamiltonians, so that the search is expressed directly in terms of pressure, strain, or chemical composition, or by targeting programmable quantum simulators, such as ultracold atoms in optical lattices~\cite{das2024realizing}, where the Hamiltonian parameters are directly tunable and the designed models could be realized and probed experimentally~\cite{kokail2026inverse}.

Our results show that machine-learning approaches such as NQS can extend variational many-body methods from the study of given Hamiltonians to the design of new ones, providing a scalable route towards the discovery of quantum matter with targeted properties.

\section{Data Availability}
The numerical simulations have been performed using the software library NetKet~\cite{netket2,vicentini2022netket}.
The code and data required to reproduce the results of this work are openly available at~\cite{repo}.

\acknowledgments
AS is supported by the Google PhD Fellowship 2025.
LLV is supported by SEFRI under Grant No. MB22.00051 (NEQS - Neural Quantum Simulation).
We acknowledge the CINECA award under the ISCRA initiative, for the availability of high-performance computing resources and support.
The Flatiron Institute is a division of the Simons Foundation.

\appendix

\section{Sign Structure of the Variational Wave Function}
\label{sec:sign}
A well-known difficulty of variational calculations on frustrated magnets is that the ground-state wave function possesses a complicated sign structure.
The difficulty is compounded in FNQS calculations, where a single ansatz must approximate the ground states of many Hamiltonians at once, and these can have markedly different sign structures.
A clear example is the $J_1$-$J_2$ Heisenberg model of~\cref{eq:j1j2_model}, whose ground-state sign structure is well described by the Marshall rule for $J_2/J_1 \lesssim 0.4$ and a collinear rule for $J_2/J_1 \gtrsim 0.6$~\cite{marshall1955antiferromagnetism}: an FNQS optimized over the full range $0 \leq J_2/J_1 \leq 1$ must reproduce both simultaneously.
Restricting the optimization to a hypercube in coupling space, as done in the inverse design framework (see~\cref{fig:artistic_figure}), reduces the number of sign structures the network has to accommodate, but does not eliminate the problem, since a trajectory may still cross a region where the sign structure changes.

A good prior for the sign structure of the ground state is provided by the solution of the corresponding classical spin model.
Starting from a generic quantum Heisenberg Hamiltonian $\hat H = \sum_{\boldsymbol{r}, \boldsymbol{r^\prime}} J_{\boldsymbol{r} \boldsymbol{r^\prime}}\, \Sv_{\boldsymbol{r}}\cdot\Sv_{\boldsymbol{r}^\prime}$, we replace the
spin operators by classical vectors $\boldsymbol{S}_{\boldsymbol{r}}$, obtaining the classical model
\begin{equation}
\label{eq:classical_hamiltonian}
H = \sum_{\boldsymbol{r}, \boldsymbol{r^\prime}} J_{\boldsymbol{r} \boldsymbol{r^\prime}}\, \boldsymbol{S}_{\boldsymbol{r}} \cdot \boldsymbol{S}_{\boldsymbol{r}^\prime}  .
\end{equation}

For the translationally invariant systems considered here, the coupling $J_{\boldsymbol{r} \boldsymbol{r^\prime}}$ depends only on the relative displacement $\boldsymbol{\delta} = \boldsymbol{r} - \boldsymbol{r^\prime}$, namely $J_{\boldsymbol{r} \boldsymbol{r^\prime}}=J_{\boldsymbol{\delta}}$.
The classical spin configuration that minimizes the energy in~\cref{eq:classical_hamiltonian} can then be determined using the well-known Luttinger-Tisza (LT) method~\cite{luttinger1946theory}.
The approach consists of writing the Hamiltonian in Fourier space, where it takes the quadratic form
\begin{equation}
    H = \sum_{\boldsymbol{q}} J_{\boldsymbol{q}}\, \boldsymbol{S}_{\boldsymbol{q}} \cdot
    \boldsymbol{S}_{-\boldsymbol{q}} ,
    \label{eq:classical_hamiltonian_fourier}
\end{equation}
with $J_{\boldsymbol{q}} = \sum_{\boldsymbol{\delta}} J_{\boldsymbol{\delta}}\, e^{i \boldsymbol{q} \cdot \boldsymbol{\delta}}$ and $\boldsymbol{S}_{\boldsymbol{q}} = \sum_{\boldsymbol{r}} \boldsymbol{S}_{\boldsymbol{r}}\, e^{i \boldsymbol{q} \cdot \boldsymbol{r}} / \sqrt{N}$ the Fourier transforms of the couplings and of the spin variables.
The minimum of~\cref{eq:classical_hamiltonian_fourier} is attained at the wave vector
\begin{equation}
    \boldsymbol Q = \arg\min_{\boldsymbol q \in \mathrm{BZ}}\, J_{\boldsymbol{q}} ,
    \label{eq:lt}
\end{equation}
which, on a finite lattice, can be determined by a grid search over the Brillouin zone (BZ).
For a Bravais lattice, a corresponding minimum-energy classical configuration is the coplanar spiral $\boldsymbol{S}_{\boldsymbol{r}} \propto \textbf{e}_1 \cos (\boldsymbol{Q} \cdot \boldsymbol{r}) + \textbf{e}_2 \sin (\boldsymbol{Q} \cdot \boldsymbol{r})$, where $\textbf{e}_1, \textbf{e}_2$ are orthonormal vectors in spin space.

Motivated by the procedure above, we define the sign rule for a many-spin configuration $\boldsymbol{\sigma} = \{\sigma_{\boldsymbol{r}}\}$, with components $\sigma_{\boldsymbol{r}} = \pm \frac{1}{2}$, as
\begin{equation}
\Phi(\boldsymbol{\sigma}) = \exp\!\Big(i \sum_{\boldsymbol{r}} \phi_{\boldsymbol{r}} \, \sigma_{\boldsymbol{r}} \Big),
\label{eq:classical_phase}
\end{equation}
where $\phi_{\boldsymbol{r}} = \boldsymbol Q\cdot\boldsymbol r \ (\mathrm{mod}\ 2\pi)$.
This phase term reproduces the sign rules known in the limiting cases of the families of Hamiltonians considered in this work.
For Néel order on the square lattice, $\boldsymbol Q=(\pi,\pi)$, yielding $\phi_{\boldsymbol{r}}=0$ on one sublattice and $\phi_{\boldsymbol{r}}=\pi$ on the other, so that~\cref{eq:classical_phase} reduces, up to a global phase, to the Marshall sign rule $\Phi(\boldsymbol{\sigma}) = \exp(i\pi\sum_{\boldsymbol{r}\in B} \sigma_{\boldsymbol{r}})$, with $B$ one of the two sublattices.
Collinear stripe order follows analogously from $\boldsymbol Q = (\pi,0)$ or $(0, \pi)$, with $B$ now every other column or row of the lattice~\cite{marshall1955antiferromagnetism}.
For the $120^\circ$ Néel order on the triangular lattice, the ordering wave vector is $\boldsymbol{Q}=\pm(2\pi/3,2\pi/3)$, yielding $\phi_{\boldsymbol{r}}=0,\pm 2\pi/3$ depending on which of the three sublattices site $\boldsymbol{r}$ belongs to.
Therefore,~\cref{eq:classical_phase} reduces to the known classical phase of the triangular antiferromagnet, $\Phi(\boldsymbol{\sigma}) = \exp[\,i\tfrac{2\pi}{3} (\sum_{\boldsymbol{r} \in B} \sigma_{\boldsymbol{r}} - \sum_{\boldsymbol{r} \in C} \sigma_{\boldsymbol{r}})\,]$, where $B$ and $C$ are two of the three sublattices~\cite{huse1988simple}.
The phase term~\cref{eq:classical_phase} can be generalized to lattices with multiple sites per unit cell, where $J_{\boldsymbol{q}}$ becomes a matrix in the sublattice indices, as for the model with eight free couplings studied in~\cref{sec:eight_coupling}.
In this case, the phases $\phi_{\boldsymbol{r}}$ are determined by the wave vector $\boldsymbol{Q}$ that minimizes the lowest eigenvalue of the matrix and by the corresponding eigenvector, the former fixing the relative phases between different unit cells and the latter those between sites within the same unit cell.

The amplitude of the variational state used in our calculations is then written as $\Psi_\theta(\boldsymbol{\sigma}|\J) = \Phi(\boldsymbol{\sigma}|\J)\, \tilde{\Psi}_\theta(\boldsymbol{\sigma}|\J)$, with $\tilde{\Psi}_\theta(\boldsymbol{\sigma}|\J)$ the output of the transformer neural network and $\Phi(\boldsymbol{\sigma}|\J)$ the term in~\cref{eq:classical_phase} evaluated with the phases $\phi_{\boldsymbol{r}}$ of the model with couplings $\J$.
We stress that the $\Phi(\boldsymbol{\sigma}|\J)$ term acts as a bias at initialization, not a constraint: its role is to absorb the part of the phase already fixed by the classical physics, leaving the neural network to learn only the corrections.
The network output $\tilde\Psi_\theta(\boldsymbol{\sigma}|\J)$ carries its own sign structure, so the optimization remains free to modify the phase of any configuration and to depart from the classical prescription wherever the true ground state requires it.
For this reason, when solving the LT minimization problem in~\cref{eq:lt}, we can also consider solutions that do not satisfy the local constraint of fixed spin length.

\bibliography{biblio}

\end{document}